\documentclass{osa-article}

\journal{osajournal}
\usepackage{siunitx}

\setprjcopyright

\articletype{Research Article}
\begin{document}

\title{Achieving Full Grating-Lobe-Free Field-of-View with Low-Complexity Co-prime Photonic Beamforming Transceivers}
\author{Aroutin Khachaturian\authormark{*}, Reza Fatemi, and Ali Hajimiri}
\address{California Institute of Technology, USA\\}
\email{\authormark{*}akhachat@caltech.edu} %% email address is required

\begin{abstract}

Integrated photonic active beamforming can significantly reduce the size and cost of coherent imagers for LiDAR and medical imaging applications. In current architectures, the complexity of photonic and electronic circuitry linearly increases with the desired imaging resolution. We propose a novel photonic transceiver architecture based on co-prime sampling techniques that breaks this trade-off and achieves the full (radiating-element-limited) field-of-view (FOV) for a 2D aperture with a single-frequency laser. Using only order-of-$N$ radiating elements, this architecture achieves beamwidth and side-lobe level (SLL) performance equivalent to a transceiver with order-of-$N^2$ elements with half-wavelength spacing. Furthermore, we incorporate a pulse amplitude modulation (PAM) row-column drive methodology to reduce the number of required electrical drivers for this architecture from order of $N$ to order of $\sqrt{N}$. A silicon photonics implementation of this architecture using two 64-element apertures, one for transmitting and one for receiving, requires only 34 PAM electrical drivers and achieves a transceiver SLL of \SI{-11.3}{\dB} with 1026 total resolvable spots, and $0.6^\circ$ beamwidth within a $23^\circ \times 16.3^\circ$ FOV.
\end{abstract}

%%%%%%%%%%%%%%%%%%%%%%%%%%  body  %%%%%%%%%%%%%%%%%%%%%%%%%%
\section{Introduction}     %%%%%%%%%%%%%%%%%%%%%%%%%%  %%%%%%%%%%%%%%%%%%%%%%%%%%  %%%%%%%%%%%%%%%%%%%%%%%%%%

Solid-state photonic platforms provide an integration pathway towards for many photonics applications ranging from communications and medical imaging \cite{Eggleston:18} to inertia sensors \cite{Parham:Gyro} and LiDAR imagers \cite{Aflatouni_NI:15, Fatemi:17, White2020}. In particular, integrated active beamformers, also known as optical phased arrays (OPAs), implemented in silicon photonic platforms have the potential to perform complex and high-speed wavefront manipulation and processing on a single mass-producible chip\cite{Fatemi:17, Poulton:19, Miller:18, Watts:Nature}, which can outperform their bulk optics and MEMS counterparts \cite{TUANTRANONT:01, Wang:19}. This can lead to lens-free, miniaturized, and low-cost coherent imaging systems with applications in LiDAR scanners, robotics, bio-medical imaging, optical communication, and remote sensing. Early efforts in the past decade have focused on demonstrating different architectures for chip-scale photonic beamforming systems \cite{Watts:Nature, Abediasl:2017, Aflatouni:15, Doylend:11}. In these systems, implementation complexity significantly increases from the photonic front-end for wavefront manipulation and processing to the back-end for electrical processing as the number of resolvable spots scales. As the number of pixels increases, the required number of photonic radiators, phase shifters, and electrical interconnect nodes grows and the overall power consumption, form factor, and cost can become prohibitive. For instance, to address the interconnect density challenge, complex and costly electrical interconnect solutions such as through-silicon via (TSV) \cite{Poulton:20}, monolithic platforms \cite{Chung:17}, or large-scale chip-to-chip interposers are investigated \cite{Kim:19}. Alternatively, full wavefront control can be sacrificed with architectures that can only do simple beamforming to reduce the interconnect density challenges \cite{Ashtiani:19,Chung:17,Dostart:20}.
\par

\begin{figure}[!t]
	\includegraphics[width=\linewidth]{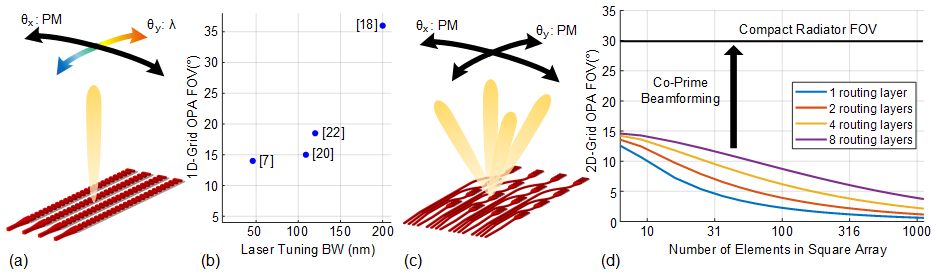}
	\caption{Solid-state beam steering methods. (a) 1D-grid aperture beam steering with a tunable laser source. (b) FOV of 1D apertures as a function of wavelength tuning range for prior art \cite{Watts:512,Dostart:20,Miller:18,Kim:2019}. (c) 2D-grid aperture beam steering with a fixed-wavelength laser. (d) FOV of 2D-grid uniform apertures as a function of the number elements in the array for a different number of photonics routing layers. 2D-grid co-prime transceiver OPAs can operate in a radiating-element-limited FOV regime using a single frequency source. }
	\label{1Dvs2D}
\end{figure}

These scalability bottlenecks are the direct result of the OPA architectural choices. The most common solid-state 2D steerable beamformers primarily utilize OPAs with 1D apertures \cite{Poulton:19, Miller:18, Abediasl:2017, Dostart:20, Poulton:17}, as shown in Fig. \ref{1Dvs2D}(a). These OPAs rely on long wavelength-sensitive grating-based radiators in conjunction with a widely tunable integrated laser to steer the beam in one direction by around $20^\circ$ \cite{Watts:512, Ma:20, Dostart:20} and steer in the perpendicular direction via phase tuning. These architectures require a rapidly tunable laser. These architecture topically require rapid wavelength tuning over about \SI{100}{\nano \meter} of wavelength (dictated by the beam scanning rates), resulting high-complexity and high-cost widely tunable laser sources. Furthermore, 1D OPAs cannot perform additional complex wavefront processing along the direction steered by the wavelength. Fig. \ref{1Dvs2D}(b) shows the laser wavelength tuning range required to achieve the desired FOV for several 1D OPA implementations. An important performance metric is the complexity order of the system, defined as the number of integrated components required compared to the number of resolvable spots. As a baseline, 1D OPAs, which require $N$ phase shifters and $N$ radiators for $N$ resolvable spots, have a complexity of order $N$.\par

On the other hand, 2D apertures, as shown in Fig. \ref{1Dvs2D}(c), can operate with a low-cost single-wavelength laser. Moreover, 2D OPAs can, in principle, surpass the beam steering limitation imposed by a tunable laser's finite wavelength-sweep range, and fully phase-controlled 2D OPAs can generate arbitrary wavefronts; however, they require a 2D grid of radiators and the corresponding phase shifters, and hence the system complexity is on the order of $N^2$. In addition to this increased system complexity, 2D apertures suffer from planar integrated photonic routing limitations, resulting in a limited effective field-of-view \cite{Fatemi:17}. In this scenario, routing photonics waveguides to inner elements of the array requires a large pitch between the individual radiating elements. Furthermore, this pitch itself has to increase with increasing number of elements, resulting in very poor scaling to large arrays. For planar photonics platforms with dielectric waveguides and radiators, this pitch is greater than half the wavelength. For arrays with larger than half-wavelength element spacing, the angular spacing of the grating lobes is calculated using 
\begin{equation}
    \theta_{GL}=sin^{-1}(\lambda/d_x)
    \label{GL}
\end{equation}
where $d_x$ is the pitch of the radiating elements and $\lambda$ is the wavelength \cite{HansenPA}. This effectively limits the useful FOV of the aperture to the angular spacing of two grating-lobes in an array.\par

While multi-layer photonics platforms \cite{Poon:2017} may alleviate this problem to a limited degree, they do not come anywhere close to solving it. Fig. \ref{1Dvs2D}(d) shows the trade-off between the number of radiators in an array and the effective grating-lobe-limited FOV for different numbers of photonic routing layers. For this plot, the radiating element and the output waveguide pitch are \SI{2}{\micro \meter} and \SI{1}{\micro \meter}, respectively. One approach to ameliorate the 2D OPA routing constraint is to design a non-uniform sparse array \cite{Reza:JSSC} that can suppress the grating-lobes. The randomized positions of OPA radiating elements in a 2D grid can be optimized to achieve the desired beamwidth, SLL, as well as meet planar photonics routing constraints. Alternatively, one can use other array beamforming techniques such as vernier arrays \cite{Dostart:19} to relax photonic routing limitations. However, the system complexity order of a 2D-grid aperture remains a challenge.\par

\par

This work addresses the 2D-grid aperture OPA system complexity challenge using co-prime sampling techniques in uniform arrays with order-of-$N$ system complexity instead of order-of-$N^2$ complexity. We demonstrate this approach using a novel transceiver OPA architecture with a 2D-grid aperture that resolves the 2D routing constraints and simplifies implementation complexity. This transceiver architecture results in an effective FOV primarily limited by the individual radiating element pattern. It operates with a single-wavelength laser while maintaining an order-of-$N$ system complexity, similar to its 1D-grid OPA counterparts. This is achieved by co-designing the transmitter and receiver apertures with co-prime radiating element spacing, which leads to co-prime angular beam spacing that overlaps only in a single direction. Furthermore, we incorporate a row-column driver configuration that further reduces electronic drive circuitry complexity to the order of $\sqrt{N}$. We present the implementation of such a co-prime transceiver OPA in a standard silicon photonics process achieving a full (radiator-limited) FOV and 1026 resolvable spots. This 2D-grid co-prime transceiver architecture with two $8\times8$ transmitter and receiver apertures, a fixed-wavelength laser source, and 34 electrical PAM drives, achieves similar angular resolution to a 32-element 1D-grid OPA with a tunable laser, or a $32\times32$-element 2D-grid OPA with a fixed-wavelength laser.\par

\section{Co-Prime Optical Beamforming}   %%%%%%%%%%%%%%%%%%%%%%%%%%  %%%%%%%%%%%%%%%%%%%%%%%%%%  %%%%%%%%%%%%%%%%%%%%%%%%%%

\begin{figure}[!t]
	\includegraphics[width=\linewidth]{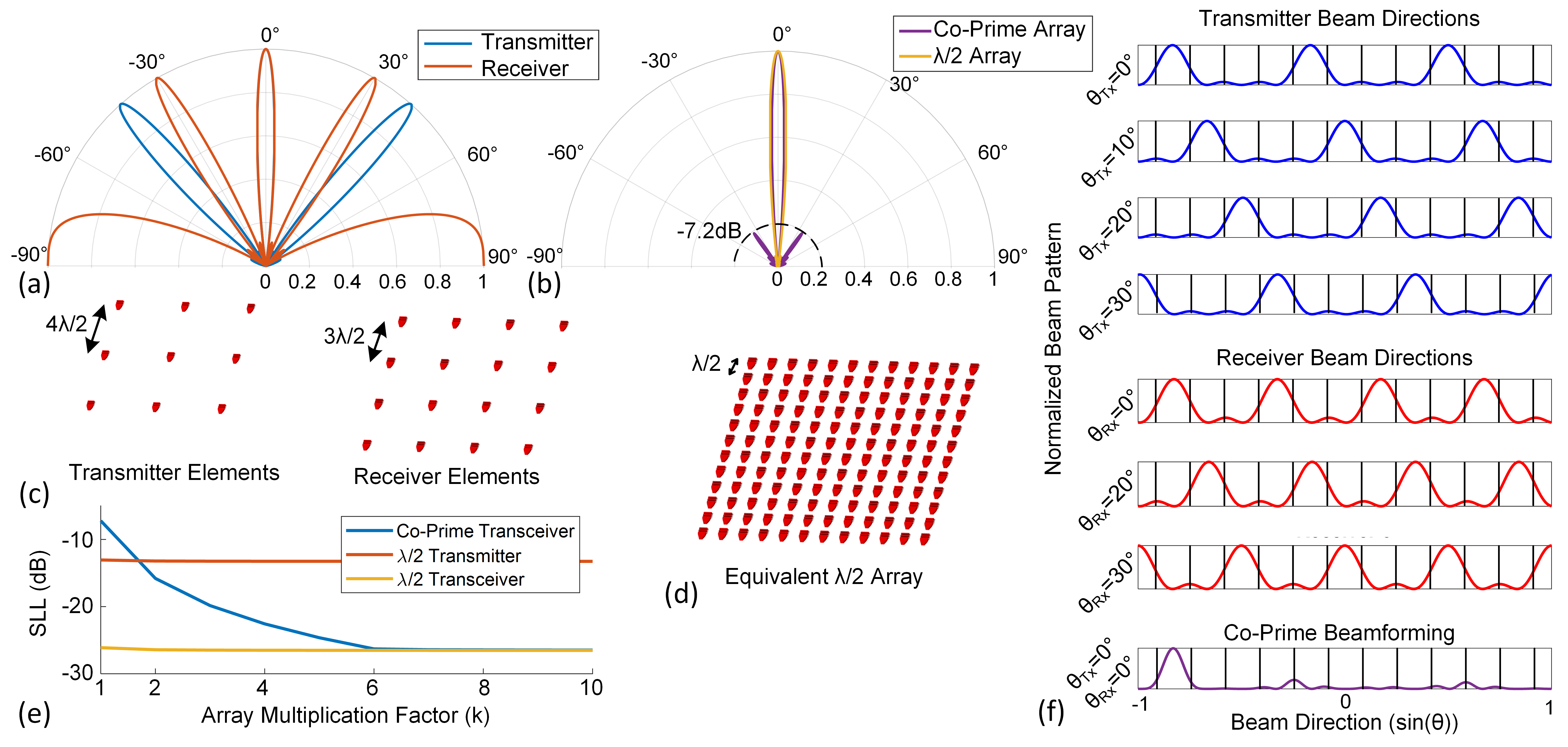}
	\caption{Co-prime beamforming example for $P=3$, $Q=4$. (a) Cross-section of the co-prime transmitter and receiver far-field radiation patterns. (b) Cross-section of the transceiver co-prime beam showing no grating lobes and the equivalent half-wavelength spacing array far-field radiation pattern. (c) Co-prime transmitter and receiver array elements on larger than half-wavelength spacing grid. (d) Half-wavelength spacing array with beamwidth equivalent to the co-prime array. (e) Transceiver SLL as a function of array multiplication k. (f) Co-prime array beam steering for several directions and the resulting transceiver pattern for $\theta_{Rx}=0^\circ$ and $\theta_{Tx}=0^\circ$.}
	\label{CPSampling}
\end{figure}

The presented photonic co-prime phased array utilizes co-prime sampling \cite{PPV:2010, PPV:2011} to synthesize a transceiver pattern with no grating lobes (no aliasing) using 2D-grid transmitter and receiver OPAs with larger than half-wavelength spacing. A phased array transceiver with independent transmitter and receiver apertures, co-located in close proximity to each other, has an overall transceiver beam pattern given by the product of the transmitter and receiver array factors
\begin{equation}
	 P_{TRx}(\theta,\theta_{Rx},\theta_{Tx}) = P_{Rx}(\theta,\theta_{Rx})\cdot P_{Tx}(\theta,\theta_{Tx}) \cdot \rho(\theta),
	\label{E_TRX} 
\end{equation}
where $\theta_{RX}$ and $\theta_{TX}$ are beam steering directions for the two arrays, and $\rho$ is the reflection coefficient of the imaging target. If the transmitter and receiver element spacings ($d_{TX}$ and $d_{RX}$) are defined to be co-prime integers within a constant factor of each other, $d_{TX}=Pd_x$ and $d_{RX}=Qd_x$, where $P$ and $Q$ are co-prime integers, the transmitter radiates in several directions, and the receiver receives from several directions; however, only one of those received directions overlaps with one of the transmitted directions.  Furthermore, the transceiver array will have a synthesized pattern with grating lobes equivalent to two uniform transmitter and receiver arrays with $d_x$ radiating element pitch. If $d_x=\lambda/2$, then the synthesized transceiver will have no grating-lobes in the FOV. In other words, co-prime spacing of the transmitter and receiver elements suppresses the grating lobes and enables two-dimensional beam steering over the full FOV limited by the radiation pattern of individual elements. 
\par

Fig. \ref{CPSampling} shows an example of co-prime beamforming using co-prime integers $P=3$ and $Q=4$. The transmitter array contains $N_{Tx}=4$ elements with $d_{Tx}=3\lambda/2$ spacing and the receiver is comprised of $N_{Rx}=3$ elements with $d_{Rx}=4\lambda/2$. While the transmitter and receiver arrays contain three and four grating-lobes respectively (Fig. \ref{CPSampling}(a)), the synthesized transceiver pattern shown in Fig. \ref{CPSampling}(b) contains no grating-lobes. The transmitter and receiver beam can be steered to resolve all the pixels within the FOV. Fig \ref{CPSampling}(f) shows several transmitter and receiver beam steering configurations. Combining any of the transmitter and receiver beams directions will have no grating lobes.

The number of transmitter and receiver elements can be increased to $N_{Tx}=k_1Q$ and $N_{Rx}=k_2P$ with $k_1,k_2>1$ to reduce SLL. Fig. \ref{CPSampling}(e) shows the relationship between SLL and the common array size multiplication factor $k=k_1=k_2$ for the array in Fig. \ref{CPSampling}(c). It is clear that $k=2$ is sufficient for the co-prime transceiver OPA to surpass the SLL of uniform transmitter OPA, and $k=6$ is sufficient for the co-prime transceiver OPA SLL to reach those of a half-wavelength spacing transceiver OPA.

The 2D-grid OPA in Fig. \ref{CPSampling}(c) with $k=1$ has a beamwidth of $6.3^\circ$ and can resolve $748$ points. This transceiver contains $Q^2=16$ transmitter radiators and $P^2=9$ receiver radiators. A half-wavelength spacing transmitter OPA with the same number of spots, as shown in Fig. \ref{CPSampling}(d), will require $(PQ)^2=144$ radiating elements and phase shifters to achieve the same beamwidth as the co-prime variant (Fig. \ref{CPSampling}(b)). For a factor $N$ defined as $N=PQ$, the co-prime OPAs with $N$ elements achieve similar beamwidth performance to a uniform $N^2$ element OPAs with half-wavelength element spacing. Hence, the system complexity is reduced from the order of $N^2$ to the order of $N$.

\section{Design}  %%%%%%%%%%%%%%%%%%%%%%%%%%  %%%%%%%%%%%%%%%%%%%%%%%%%%  %%%%%%%%%%%%%%%%%%%%%%%%%%

\begin{figure}[t!]
	\includegraphics[width=\linewidth]{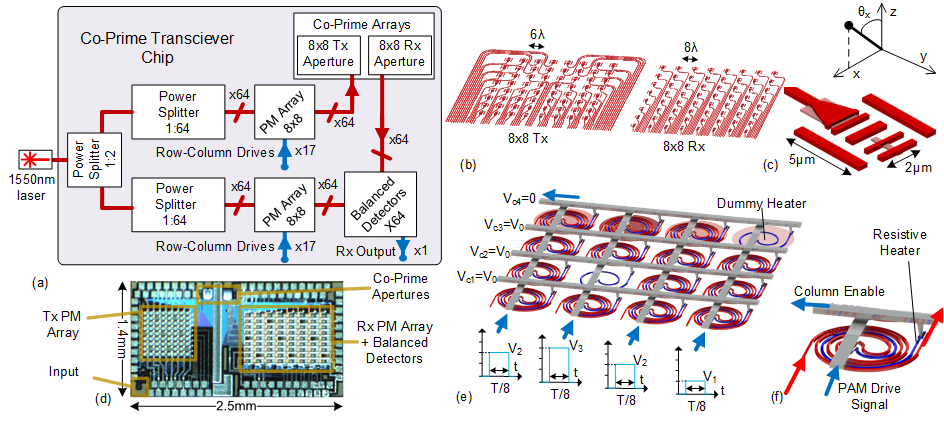}
	\caption{Co-Prime transceiver system architecture. (a) Block diagram of the co-prime transceiver. (b) Transmitter and receiver apertures implementations. (c) Compact radiator design., (d) Die photo of the fabricated chip. (e) Row-column drive phase modulator (PM) array. (f) Compact spiral thermal phase shifter.}
	\label{OPA1_BD}
\end{figure}

A silicon photonics co-prime transceiver was designed and implemented using the Advanced Micro Foundry's (AMF) standard photonics foundry to demonstrate co-prime beamforming capability. Fig. \ref{OPA1_BD}(a) shows the block diagram of the chip. Coupled power into the chip splits equally between the transmitter block for illumination beamforming and the receiver array for receiver beamforming and heterodyne detection \cite{Fatemi:17}. An $8\times8$ array of transmitter elements and an $8\times8$ array of receiver elements (Fig. \ref{OPA1_BD}(b)) with equal power distribution are used. The equivalent uniform array pitch is $d_x=2\lambda$ which results in a usable FOV of $30^\circ$. The co-prime numbers are $P=3$ and $Q=4$. This results in a transmitter array with $6\lambda=\SI{9.2}{\micro \meter}$ element pitch (grating lobes at $9.55^\circ$) and a receiver array with $8\lambda=\SI{12.4}{\micro \meter}$ element pitch (grating lobes at $7.18^\circ$) for an operational wavelength of \SI{1550}{\nano \meter}. Array multiplication factors are $k_1=2$ and $k_2=2.67$ which means the ideal SLL should be better than \SI{15}{\decibel} with $0.65^\circ$ beamwidth.

\par

A compact \SI{2}{\micro \meter} $\times$ \SI{5}{\micro \meter} radiating element was optimized and implemented as the transmitting and receiving element (Fig. \ref{OPA1_BD}(c)). This compact radiator has a \SI{3}{\decibel} far-field beamwidth of $23^\circ \times 16.3^\circ$ which becomes the FOV-limiting factor in this design. The \SI{1}{\decibel} spectral bandwidth of the radiator is over \SI{400}{nm}, making the radiators very robust to changes in the operating wavelength. At the operating wavelength of \SI{1550}{nm}, the peak radiation efficiency is at $\theta_y=7.4^\circ$.

\par
Both transmitter and receiver OPAs contain an $8\times8$ array of phase modulators for complete relative phase control between radiator elements in each block. A cascade of y-junctions divides the power equally among 64 radiating elements. A compact spiral thermo-optic phase shifter is designed for improved modulation efficiency and reduced cross-talk as shown in Fig. \ref{OPA1_BD}(f). This spiral thermo-optic phase shifter has been previously characterized in \cite{Reza:JSSC} with \SI{21.2}{\milli \watt} required electrical power for $2\pi$ phase shift and $19kHz$ electro-optic modulation bandwidth. In addition, a series of dummy thermal heaters are distributed across the thermo-optic phase shifters to compensate for the temperature gradients on-chip. The phase shifters are connected in a row-column grid (Fig.\ref{OPA1_BD}(e)), resulting in a total of $34$ electrical connections.

These row and column nodes are driven in a row-column fashion using time-domain demultiplexing \cite{Reza:JSSC}. $17$ pulse amplitude modulators (PAM) drivers continuously program the thermo-optic phase shifters where each column is active $1/8$ of the cycle $T$ and receives $8$ times the required power. Given the kilohertz range bandwidth of the modulators, cycling through the columns at megahertz frequencies ($T=\SI{4}{\mega \hertz}$) ensure that the phase shifters receive constant electrical power. This row-column modulation reduces system interconnect and drive complexity by allowing $2N+1$ drivers (+1 for the additional dummy heaters) to control $N^2$ thermo-optic phase shifters independently and hence reduce the interconnect complexity from order $N$ to order $\sqrt{N}$ without sacrificing complex beamforming capability.\par

The receiver array is configured as a heterodyne receiver with LO path phase-shifting for improved receiver SNR \cite{Fatemi:17}. The electrical output of all balanced detectors is combined on-chip to benefit from the array gain factor and boost SNR in the output signal prior to off-chip amplification and detection. The entire design is \SI{2.5}{\milli \meter}$\times$\SI{1.4}{\milli \meter} as shown in Fig. \ref{OPA1_BD}(d).

\section{Measurements} %%%%%%%%%%%%%%%%%%%%%%%%%%  %%%%%%%%%%%%%%%%%%%%%%%%%%  %%%%%%%%%%%%%%%%%%%%%%%%%%

\begin{figure} [!t]
	\includegraphics[width=\linewidth]{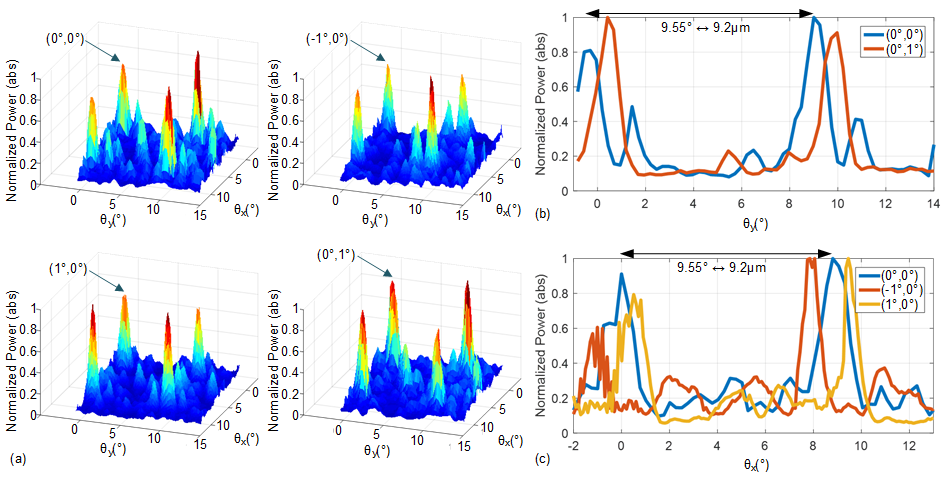}
	\caption{Co-prime transmitter beamforming and steering. Grating lobes are spaced 9.55$^\circ$ consistent with $9.2\mu m$ spacing of radiating elements. (a) 2D optimized beam pattern for four directions. (b) Cross-section of $\theta_y$ plane. (c) Cross-section of $\theta_x$ plane.}
	\label{OPA1_TxPatt}
\end{figure}

\begin{figure} [!t]
	\includegraphics[width=\linewidth]{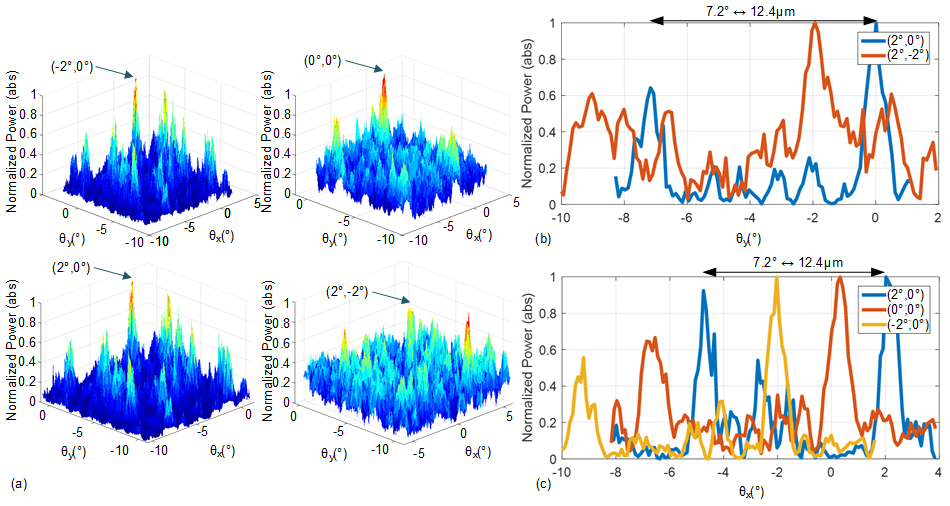}
	\caption{Co-prime receiver beamforming and steering. Grating lobes are spaced 7.2$^\circ$ consistent with $12.4\mu m$ spacing of radiating elements. (a) 2D optimized beam pattern for four directions. (b) Cross-section of $\theta_y$ plane. (c) Cross-section of $\theta_x$ plane.}
	\label{OPA1_RxPatt}
\end{figure}

A photonic far-field pattern measurement setup was constructed for the transceiver. This setup allows for independent transmitter and receiver optimization with high sensitivity. The transmitter beam pattern is measured by scanning the far-field radiated power from the chip (modulated at \SI{1.2}{\mega \hertz}) in the $\theta_x$ and $\theta_y$ directions. Far-field radiated power was collected by an InGaAs photodetector. The transmitter beam pattern was optimized for several points demonstrating 2D beam steering capability in both directions. Fig. \ref{OPA1_TxPatt}(a) demonstrates four examples of 2D beam patterns measured in four directions with clear grating lobes visible at around $9.5^\circ$ spacing in both directions as expected. The cross-sections of these beam pattern in $\theta_x$ and $\theta_y$ directions are shown in Fig. \ref{OPA1_TxPatt}(b-c). Subsequently, the receiver array is characterized by illuminating the receiver aperture using a cleaved fiber. To remove the random phase fluctuations between the illumination and LO paths \cite{Fatemi:17}, the input light was externally modulated using two SSB modulators at \SI{10}{\mega \hertz} and \SI{11.5}{\mega \hertz} respectively. The mixed downconverted signal at \SI{1.5}{\mega \hertz} was amplified off-chip for processing. This setup is used to optimize the receiver beam in several directions. Four such patterns are shown in Fig. \ref{OPA1_RxPatt}(a). The grating lobes are visible at $7.2^\circ$ in both directions, consistent with the design of the OPA. Cross-sectional view of these beam patterns for several directions are shown in Fig. \ref{OPA1_RxPatt}(b-c).

\begin{figure} [!b]
	\includegraphics[width=\linewidth]{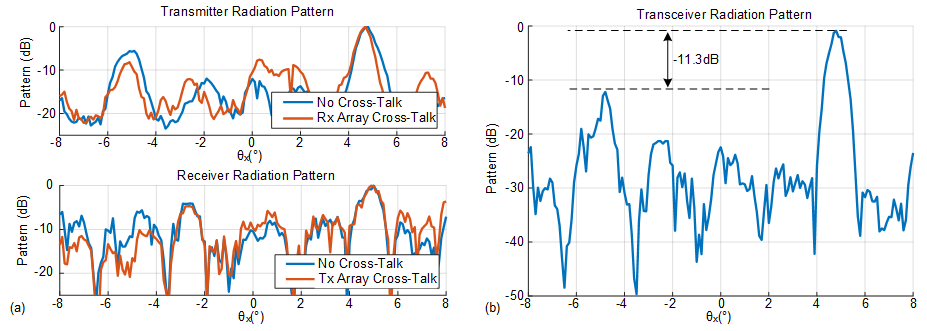}
	\caption{Overlap plot of the transmitter and receiver patterns. (a) Beam patterns captured when the optimized settings are loaded separately (blue) and when both settings are loaded concurrently (red). (b) Synthesized transceiver pattern.}
	\label{OPA1_TxRx_1D}
\end{figure}

%Finally,
Demonstrating the 2D beam steering capability of the transmitter and the receiver array ensures that the main beam of the two apertures can be co-aligned in the same direction for all points in the 2D FOV. For a given resolvable spot (pixel), the transmitter and receiver array can be simultaneously co-aligned on that point, and the co-prime nature of the transceiver will limit in the receiver aperture to collect signal from that particular direction and suppress all the signals from the grating lobes of the transmitter. To demonstrate co-prime grating-lobe suppression capability of the transceiver, the full system was characterized with concurrently active transmitter and receiver arrays. Blue patterns in Fig. \ref{OPA1_TxRx_1D}(a) show a 1D measurement of the formed transmitter and receiver beam over a $16^\circ$ range, displaying the expected grating lobes. Programming the two phased arrays simultaneously (orange patterns in Fig. \ref{OPA1_TxRx_1D}(a)) showed that the thermal cross-talk between the two patterns causes less than \SI{0.5}{\decibel} disturbance in the main beam power of the transmitter and receiver with a worst-case of \SI{5}{\decibel} increased side-lobe level for the transmitter array. The combined synthesized pattern for the transceiver array is calculated from the beam patterns and shown in Fig. \ref{OPA1_TxRx_1D}(b). For the full scan range of $16^\circ$, the highest side-lobe level is at \SI{-11.3}{\decibel} with a transceiver beamwidth of $0.6^\circ$ which is in close agreement with the simulations. 

\section{Discussion} %%%%%%%%%%%%%%%%%%%%%%%%%%  %%%%%%%%%%%%%%%%%%%%%%%%%%  %%%%%%%%%%%%%%%%%%%%%%%%%%

\par
The proof-of-concept implementation presented here demonstrated the realization of a co-prime transceiver architecture, achieving 1026 resolvable spots using only 128 radiating elements with only 34 electrical drivers. The advantage of this architecture is more significant for larger arrays. For example, when the number of desired pixels is $10,000$ points, the total number of required radiators and phase shifters pairs is $200$, and the total number of electrical driver circuits required is $40$. This is more than an order of magnitude of reduction in the complexity of the photonic front-end and more than two orders of magnitude reduction in the number of interconnects and electrical drivers. It is worth noting that the chip area is dominated by the phase modulator, as shown in Fig. \ref{OPA1_BD}(d). Even if half-wavelength spacing of radiators was possible, the chip area for an equivalent half-wavelength spacing OPA would be an order of magnitude larger  than the co-prime array. These characteristics make the co-prime transceiver architecture a very promising candidate for silicon photonic beamforming and coherent imaging applications due to the lower complexity, while achieving high resolution, low cost, and low power consumption.
\par

\section*{Acknowledgments}
The authors would like to acknowledge Behrooz Abiri and Parham Porsandeh Khial for their valuable inputs in the design and analysis of this work.

\section*{Disclosures}

The authors disclose no conflict of interest.

\bibliography{sample}

\end{document}